\documentclass{article}
\usepackage[utf8]{inputenc}
\usepackage{textcomp}
\usepackage{amsmath}
\usepackage{amssymb}
\usepackage{graphicx}
\usepackage{braket}
\usepackage{url} 
\usepackage{comment} 
\usepackage{subfigure}
\usepackage[makeroom]{cancel}
\usepackage{xcolor}
\usepackage{mathtools}
\usepackage{authblk}

  \usepackage[
    backend=biber,
    style=ieee,
  ]{biblatex}

\begin{document}

\title{\Large Toward Local Madelung Mechanics in Spacetime}

\author[1,a,*]{Mordecai Waegell}

\affil[1]{\small{Institute for Quantum Studies, Chapman University, Orange, CA 92866, USA}}

\affil[a]{ORCID: 0000-0002-1292-6041}

\affil[*]{Corresponding Author, email: waegell@chapman.edu}

\date{\today}

\maketitle

\begin{abstract}
It has recently been shown that relativistic quantum theory leads to a local interpretation of quantum mechanics wherein the universal wavefunction in configuration space is entirely replaced with an ensemble of local fluid equations in spacetime.  For want of a fully relativistic quantum fluid treatment, we develop a model using the nonrelativistic Madelung equations, and obtain conditions for them to be local in spacetime.  Every particle in the Madelung fluid is equally real, and has a definite position, momentum, kinetic energy, and potential energy.  These are obtained by defining quantum momentum and kinetic energy densities for the fluid and separating the momentum into average and symmetric parts, and kinetic energy into classical kinetic and quantum potential parts.  The two types of momentum naturally give rise to a single classical kinetic energy density, which contains the expected kinetic energy, even for stationary states, and we define the reduced quantum potential as the remaining part of the quantum kinetic energy density.  We treat the quantum potential as a novel mode of internal energy storage within the fluid particles, which explains most of the nonclassical behavior of the Madelung fluid.  For example, we show that in tunneling phenomena the quantum potential negates the barrier so that nothing prevents the fluid from flowing through.  We show how energy flows and transforms in this model, and that enabling local conservation of energy requires defining a quantum potential energy current that flows through the fluid rather than only flowing with it.  The nonrelativistic treatment generally contains singularities in the velocity field, which undermines the goal of local dynamics, but we expect a proper relativistic treatment will bound the fluid particle velocities at $c$.
     \end{abstract}

\section{Introduction}

Since the early days of quantum theory, researchers like Madelung \cite{madelung1927quantum} have recognized that the Schr\"{o}dinger dynamics can be recast as equations of fluid dynamics, where the conservation of probability current is mapped to conservation of fluid current.  This leads naturally to a many-worlds interpretation of Born rule probabilities as proportions of fluid current \cite{waegell2023local}, where individual experiences of collapse are mapped to individual particles within the fluid (e.g., if we observe a quantum particle to reflect off a beam splitter with twice the probability that it is transmitted, then twice as many of its constituent fluid particles are reflected as are transmitted).

For a single quantum particle (say, the electron), this dynamics occurs in the familiar three spatial dimensions and one time dimension, but for $N$ entangled particles, the standard nonrelativistic treatment occurs in a configuration space with $3N$ spatial dimensions, although there is still only one time dimension.  The fluid analogy still holds in this space, but now a single `particle' in the fluid describes all $N$ particles in 3-space at once.  For a fluid in 3-space, we can consider the kinetic and potential energies of the individual particles in the fluid, and corresponding energy densities for the bulk fluid itself, but these classical quantities are not well-defined for the `particles' in $3N$-dimensional space, which significantly undermines the analogy.

However, the local interpretation of relativistic quantum physics from 
\cite{waegell2023local} restores all of the dynamics to 3-space, which allows the full classical analogy to be recovered (see also Appendix B).  In this treatment a single quantum particle typically comprises several different fluids in spacetime, with indexes to indicate which distinct fluid each particle belongs to.  The different fluids of a given quantum system only interact with one another when there is a local coupling to another quantum system.  The indexed fluids of a single quantum system allow for the treatment of entanglement and internal degrees of freedom like spin, with all of the dynamics occurring in 3-space where the particle energies and fluid energy densities are all well-defined.  These are essentially classical fluids from a mathematical standpoint, but they behave quite differently from any standard classical liquid or gas because of the quantum potential.  To avoid confusion with the terminology, a \textit{quantum particle} refers to a particle that we typically treat using a wavefunction and corresponding probability density, like a single electron.  When this probability density is reinterpreted as a continuum fluid density, we will also speak of the \textit{fluid particles}, which are essentially classical point particles that form the elements of the fluid.  For example, corresponding to one electron, there are essentially an infinite number of fluid particles in the continuum fluid picture, all in 3-space.

The purpose of this research program is to interpret the quantum potential as a standard classical energy in 3-space and ideally to identify the local interaction rules for the particles in the fluid.  We will begin with the nonrelativistic Schr\"{o}dinger equation and the corresponding Madelung equations, with the ultimate aim of carrying the physical intuition we develop here over to the relativistic case.  Unlike much of the literature on Madelung's equations, our goal here is not to show that the fluid picture is more fundamental and that the Schr\"{o}dinger equation emerges from it, but rather to use the single-particle Madelung equations as a pseudo-classical interpretation of the Schr\"{o}dinger equation and Born rule in 3-space.

\section{Model Overview}

We begin by defining a quantum momentum density $p_q(\vec{x},t)$ using the integrand of the expectation value of the quantum momentum operator of a single-quantum-particle Madelung fluid in 3-space, $\psi^*\hat{p}\psi$, with the wavefunction expressed in eikonal form, $\psi(\vec{x},t)  \equiv R(\vec{x},t)e^{iS(\vec{x},t)/\hbar}$.  The real part of this density is the fluid current, with particle velocity $\vec{\nabla}S/m$, and the imaginary part is associated with additional momentum that sums to zero over all of the fluid particles at a given event in spacetime, with particle velocity $\Big|\frac{\hbar}{m}\frac{\vec{\nabla}R}{R}\Big|\hat{r}_i$, and unit vector $\hat{r}_i$.  This gives us a distribution of definite particle velocities at each event, such that $\sum_i \hat{r}_i =0$, so the imaginary part contributes nothing to the net current of the fluid.

Next, we define a classical energy density using the integrand of the expectation value of the quantum kinetic energy operator, $\psi^*\hat{K}\psi$, and separate it into real positive terms that correspond to the classical kinetic energy density associated with the particle velocities found above, and other real terms that can become negative, which we identify as the (reduced) quantum potential energy density.  There are also several imaginary terms we ignore, which integrate to zero.  When the Schr\"{o}dinger equation density is expanded out, those imaginary terms belong to the continuity equation, while the real terms relate to the evolution of the fluid current.  We thus find that the `quantum kinetic energy density' is really a sum of a classical kinetic and quantum potential energy densities, which suggests that the quantum potential has a kinetic character, even though it can become negative.  This is consistent with the finding in one attempted relativistic generalization of the fluid picture \cite{poirier2012trajectory} that the direct value of the quantum potential plays a role in the physics, rather than only its derivatives.

There are two ways to group the terms that appear in the quantum kinetic energy density $k_q$, and each has its own conceptual advantages.  The standard way \cite{madelung1927quantum,bohm1952suggested} is to define a kinetic energy density $k_a = \frac{1}{2m}R^2|\vec{\nabla}S|^2$ corresponding to the net fluid flow, and the quantum potential energy density as $q = -\frac{\hbar^2}{2m}R\nabla^2R$, such that $k_q = k_a + q$.  However, the quantum potential density can be broken down into a kinetic energy density $k_s = \frac{\hbar^2}{2m}|\vec{\nabla}R|^2$ corresponding to the motion that averages to zero over the velocity distribution, and a reduced quantum potential $q_r = -\frac{\hbar^2}{4m}\nabla^2 R^2$, such that $q=k_s + q_r$, and $k_q = k_a + k_s + q_r$.  Summing over the velocity distribution, all of the cross terms cancel so that the classical kinetic energy density associated with the total velocity $\frac{1}{m}\big(\vec{\nabla}S +  \hbar\frac{\vec{\nabla}R}{R}\hat{r}_i\Big)$ of each fluid particle can be written as $k_c = k_a + k_s$, so the second way we can group terms is $k_q = k_c + q_r$.  This is probably the more physically correct grouping, since $k_a = 0$ for stationary states, while $k_s$ (and thus $k_c$) integrates to the expectation value of kinetic energy.  Furthermore, there is experimental evidence that there is motion in stationary states \cite{silverman1982relativistic}, because the laboratory-frame half-lives of muons in stationary states of muonic atoms are time-dilated in a way that appears to relate to the speed $\Big|\frac{\hbar}{m}\frac{\vec{\nabla}R}{R}\Big|$.

However, as we will show, the reduced quantum potential acts as an intermediary between the two kinetic energy terms, so relative to the average kinetic energy $k_a$, it is reasonable to think of both the symmetric kinetic energy $k_s$ and reduced quantum potential $q_r$ as a single entity, the usual quantum potential $q$, and we would use the standard grouping.  To see how the energy flows between these types, we construct the continuity equations for each of the above energy densities, and match their source/sink terms, which give us $k_s \textcolor{teal}{\leftrightarrow} q_r \textcolor{purple}{\leftrightarrow} k_a \textcolor{blue}{\leftrightarrow} u$, where $u = R^2U$ is the external potential energy density.  

This calculation also reveals that energy can only be locally conserved if the (reduced) quantum potential energy can flow between particles in the fluid, in addition to flowing with them.  We compute the necessary energy current for local energy conservation in the case that the reduced quantum potential $q_r$ flows through the fluid.  We also compute it for the case that the entire standard quantum potential $q$ flows through the fluid, which is consistent with a model in which all of $q$ is treated as an internal potential energy, but inconsistent with our model, since the symmetric kinetic energy $k_s$ should flow with the fluid, not through it.

We will show that any discontinuity in $u$ is exactly canceled by an opposite discontinuity in $q_r$, so the total potential energy density $q_r+u$ is a continuous function, and thus the two types of potential energy seem to naturally lock together.  It is also interesting to note that if we also include $k_s$ as part of the potential energy relative to the flow energy $k_a$, the potential energy density $q+u$ is (mostly) smooth, and appears to coincide more naturally with the fluid density, and physical intuition about how it flows.  We will demonstrate these features in our tunneling example, and show how the quantum potential negates the barrier, providing a simple explanation of how the otherwise-classical fluid flows through.

Finally, we have made two somewhat arbitrary assumptions in constructing this model, either of which might be subject to experimental falsification.  The first is that there is motion related to speed $\Big|\frac{\hbar}{m}\frac{\vec{\nabla}R}{R}\Big|$ even in stationary states.  The past experiments with muonic atoms were of low fidelity, but a new set of muon experiments in which the shape of stationary wavefunctions is carefully controlled could allow us to measure whether the average half-life of muons in the fluid is really consistent with this velocity distribution.  If this first assumption seems to be borne out, this would also lend support for the idea that only the reduced quantum potential is a true internal energy, which is our second assumption.  Either way, these experiments would probe an important and under-explored interface between quantum mechanics and relativity, which could help us to reconcile stationary quantum states with time dilation, even if this fluid model is falsified.  Beyond this, we believe a new generation of muon experiments would be invaluable for shedding light on the interplay between quantum mechanics and relativity, both special and general \cite{lobo2023muon}.

\section{The Formalism}

\subsection{The quantum momentum density}

We define the quantum momentum density  $\vec{p}_q(\vec{x},t)$ as $ \psi^*(\vec{x},t) \hat{p} \psi(\vec{x},t)$, in the sense the integral of this quantity over space gives the expectation value of the momentum.  Taking the eikonal form of the wavefunction, $\psi = Re^{iS/\hbar}$,

\begin{equation}
   \vec{p}_q(\vec{x},t) = \psi^*(\vec{x},t) \hat{p} \psi(\vec{x},t)  = -i\hbar R(\vec{x},t)e^{-iS(\vec{x},t)/\hbar}\vec{\nabla} R(\vec{x},t)e^{iS(\vec{x},t)/\hbar}  \label{p_int}
\end{equation}
\begin{equation}
     =-i\hbar Re^{-iS/\hbar}\big(\vec{\nabla} R + \frac{i}{\hbar}R\vec{\nabla} S\big)  e^{iS/\hbar} \nonumber
\end{equation}
\begin{equation}
     = R\big( R\vec{\nabla} S -   i\hbar\vec{\nabla} R \big)  \nonumber.
\end{equation}
We define the \textit{average momentum density} as the real part of the quantum momentum density,
\begin{equation}
    \vec{p}_a(\vec{x},t) \equiv R^2 \vec{\nabla} S = m \rho \vec{v}_a = m\vec{j} ,
\end{equation}
where the density $\rho \equiv R^2$ is understood as describing a locally conserved fluid, and the average local velocity of the fluid-particles is $\vec{v}_a(\vec{x},t) = \vec{j}(\vec{x},t)/\rho(\vec{x},t) = \vec{\nabla}S(\vec{x},t)/m$, where $\vec{j}$ is the probability (fluid) current.  

As a brief aside and cautionary note, the spatial integral over this density gives the expectation value of the momentum, which is consistent with the ensemble average after measuring the particle momentum many times, where each result appears to collapse to a random momentum eigenstate with Born rule probability.  However, unlike a position measurement, where the ensemble average probability to find the particle near $\vec{x}$ matches the density $\rho(\vec{x})$, the momentum eigenstates are spread over space.  To find the probability of each outcome one must decompose the entire fluid state into this set of delocalized states, which has little to do with the density $\vec{p}_q(\vec{x})$ at each location $\vec{x}$, so there is no joint probability distribution for which the outcomes of both position and momentum measurements are marginals (Wigner functions are as close as one can get).  In general, it is the measurement apparatus which determines the eigenbasis into which the fluid must be decomposed, and with the exception of position measurements, these eigenstates are spread out in space.  However, the local density $\vec{p}_q(\vec{x})$ is of physical importance, because it correctly describes where the momentum can be found, regardless of what basis is used to measure it.  For example, if the fluid is coherently separated into multiple regions, and local momentum modes are measured in those separate regions, the expectation value in each region is still the integral of $\vec{p}_q(\vec{x})$ over that region.

Now, the expected classical kinetic energy of a fluid particle with velocity $\vec{v}_a(\vec{x},t)$ at $\vec{x}$ and $t$ is then $K_a(\vec{x},t) = \frac{1}{2}m\vec{v}_a(\vec{x},t) \cdot \vec{v}_a(\vec{x},t)$, so the expected classical kinetic energy density associated with the average local velocity of the fluid is,
\begin{equation}
    k_a(\vec{x},t) \equiv \rho(\vec{x},t)K_a(\vec{x},t) = \frac{1}{2m}R^2(\vec{\nabla}S \cdot \vec{\nabla}S).
\end{equation}

We are making progress, but there still seems to be a discrepancy in this analysis having to do with stationary states.  The classical analog of the energy eigenstates describe particles in motion, and experiments with muonic atoms show that this motion slows the proper time of the muons relative to the lab frame, but stationary states have $\vec{v}_a = 0$, so this motion seems to be missing.  

To resolve this issue, we define the imaginary part of the quantum momentum density as the \textit{symmetric momentum density},
\begin{equation}
    \vec{p}_s(\vec{x},t) \equiv -\hbar R(\vec{x},t) \vec{\nabla} R(\vec{x},t) = m \rho \vec{v}_s(\vec{x},t),
\end{equation} 
so the \textit{symmetric velocity} is
\begin{equation}
    \vec{v}_s(\vec{x},t) = -\frac{\hbar}{m}\frac{ \vec{\nabla} R(\vec{x},t)}{R(\vec{x},t)}.
\end{equation}
The kinetic energy and kinetic energy density associated with this velocity are then,
\begin{equation}
\begin{array}{ccc}
   K_s = \frac{1}{2}m\vec{v}_s\cdot\vec{v}_s =   \frac{\hbar^2}{2m} \frac{\vec{\nabla} R\cdot \vec{\nabla} R}{R^2}, & & k_s = \rho K_s = \frac{\hbar^2}{2m}\vec{\nabla} R\cdot \vec{\nabla} R.
   \end{array}
\end{equation}
This energy is not zero for stationary states, and for simple cases like the infinite square well, all of the energy is of this type, so it seems we are on the right track to identify the motion in this state.  However, this energy is not associated with a change in the fluid density $R^2$.  To explain this we argue that, unlike a classical continuum fluid, the quantum fluid does not have a single velocity at each event in spacetime $(\vec{x},t)$, but rather a distribution of velocities,
\begin{equation}
    \vec{v}_i(\vec{x},t) = \vec{v}_a(\vec{x},t) + |\vec{v}_s(\vec{x},t)|\hat{r}_i(\vec{x},t),
\end{equation}
with unit vectors $\hat{r}_{i}(\vec{x},t)$ such that,
\begin{equation}
    \sum_{i} \hat{r}_{i}(\vec{x},t) = 0  \label{rzero}
\end{equation}
(thus symmetric).  We then have $\vec{v}_a = (\vec{v}_i)_\textrm{ave}$ because all of the  $|\vec{v}_s(\vec{x},t)|\hat{r}_i(\vec{x},t)$ components cancel and contribute nothing to the net flux of fluid particles at event $(\vec{x},t)$, and $\vec{v}_a(\vec{x},t)$ really is the average of the local velocity distribution, which gives the (net) local fluid current $\vec{j}(\vec{x},t) = \rho(\vec{x},t) \vec{v}_a(\vec{x},t).$  In this picture, the fluid particles are essentially of zero volume, and do not exert direct forces on one another.  All of their interactions are collective and occur through the quantum potential, which we will demonstrate in later sections.  Effectively, all of the fluid particles contribute to creating a collective quantum potential surface, and in turn, each of the fluid particles moves in that potential.

With this velocity, the classical kinetic energy of a particle in the fluid is $K_i(\vec{x},t) = \frac{1}{2}m\vec{v}_i(\vec{x},t)\cdot \vec{v}_i(\vec{x},t)$, and the average classical kinetic energy at event $(\vec{x},t)$ is $K_c = (K_i)_\textrm{ave} = K_a + K_s$ (the cross-terms cancel), and thus the classical kinetic energy density is, 
\begin{equation}
    k_c(\vec{x},t) = \rho(\vec{x},t)K_c(\vec{x},t) = k_a + k_s,
\end{equation}

We have inferred this classical kinetic energy, and the two terms it comprises, from the quantum momentum density, and some considerations about motion in stationary states.  Next we want to see if this kinetic energy appears in the quantum kinetic energy density.

\subsection{The quantum kinetic energy, classical kinetic energy, and quantum potential energy densities}

We define the quantum kinetic energy density $k_q(\vec{x},t)$ in standard quantum theory as the real part of $ \psi^*(\vec{x},t) \hat{K} \psi(\vec{x},t)$, in the sense the integral of this quantity over space gives the expectation value of the kinetic energy.  
\begin{equation}
   \psi^*(\vec{x},t) \hat{K} \psi(\vec{x},t)  = -\frac{\hbar^2}{2m}R(\vec{x},t)e^{-iS(\vec{x},t)/\hbar}\nabla^2 R(\vec{x},t)e^{iS(\vec{x},t)/\hbar}
\end{equation}
\begin{equation}
     =-\frac{\hbar^2}{2m}Re^{-iS/\hbar}\vec{\nabla}\cdot\big(\vec{\nabla} R + \frac{i}{\hbar}R\vec{\nabla} S\big)  e^{iS/\hbar} \nonumber
\end{equation}
\begin{equation}
     =-\frac{\hbar^2}{2m}Re^{-iS/\hbar}\bigg[ \big(    \nabla^2 R + \frac{i}{\hbar}\vec{\nabla}R\cdot\vec{\nabla} S +  \frac{i}{\hbar}R\nabla^2S\big)  e^{iS/\hbar} + \frac{i}{\hbar}\big(\vec{\nabla} R + \frac{i}{\hbar}R\vec{\nabla} S\big) \cdot (\vec{\nabla}S) e^{iS/\hbar} \bigg]\nonumber
\end{equation}
\begin{equation}
     =-\frac{\hbar^2}{2m}R\bigg(     \nabla^2 R -\frac{1}{\hbar^2}R(\vec{\nabla}S \cdot \vec{\nabla}S)+ \frac{2i}{\hbar}\vec{\nabla}R\cdot\vec{\nabla} S   +  \frac{i}{\hbar}R\nabla^2S\bigg)\nonumber.
\end{equation}
The imaginary parts always integrate to zero, so while they are related to the continuity equation, dropping them here gives a real quantum kinetic energy density,
\begin{equation}
    k_q(\vec{x},t) \equiv \textrm{Re}\big(\psi^* \hat{K} \psi\big) =   -\frac{\hbar^2}{2m}R\nabla^2 R +\frac{1}{2m}R^2(\vec{\nabla}S \cdot \vec{\nabla}S).
\end{equation}
The apparent trouble with this expression is that it can be negative, which makes it difficult to interpret as a kinetic energy density.  The resolution is that it is not, in fact, only a kinetic energy, but rather sum of a true classical kinetic energy density, which is always positive, and the quantum potential energy density, which may be negative.  

We can identify the second term of the quantum kinetic energy density above as the classical kinetic energy density $k_a$ we were expecting, which is always positive, so we will define the first as the quantum potential energy density.  From Bohm/Madelung's definition of the quantum potential energy of a single classical particle at $\vec{x}$,
\begin{equation}
    Q(\vec{x},t) \equiv -\frac{\hbar^2}{2m}\frac{\nabla^2 R}{R} = \frac{\hbar^2}{2m}\frac{\vec{\nabla} R \cdot \vec{\nabla} R}{R^2} -\frac{\hbar^2}{4m}\frac{ \nabla^2 (R^2)}{R^2} ,
\end{equation}
we see that this really is the density of quantum potential energy in the fluid of particles,
\begin{equation}
    q(\vec{x},r) \equiv  \rho(\vec{x},t) Q(\vec{x},t) = -\frac{\hbar^2}{2m}R\nabla^2 R.\label{quantpot}
\end{equation}
\begin{equation}
    = \frac{\hbar^2}{2m}\vec{\nabla} R \cdot \vec{\nabla} R -\frac{\hbar^2}{4m} \nabla^2 (R^2). \nonumber
\end{equation}
Note that in the expanded form of the quantum potential, we can now identify the first term as our other classical kinetic energy density $k_s$, so if we collect $k_s$ and $k_a$ together into a single classical kinetic energy density $k_c$, we are left with a different term that we call the \textit{reduced quantum potential} energy density,
\begin{equation}
    q_r \equiv q - k_s = -\frac{\hbar^2}{4m} \nabla^2 (R^2),
\end{equation}
corresponding to reduced quantum potential energy per fluid particle,
\begin{equation}
    Q_r = \frac{q_r}{R^2} = -\frac{\hbar^2}{4m} \frac{\nabla^2 (R^2)}{R^2}.
\end{equation}
Thus we have broken the quantum kinetic energy density down into a well-behaved positive kinetic energy density and a new reduced quantum potential which represents a new type of internal energy carried by particles in the fluid,
\begin{equation}
    k_q = k_a +  q = k_a +k_s+  q_r= k_c +  q_r.
\end{equation}
Defining the external potential energy density as $u(\vec{x},t) \equiv  \rho(\vec{x},t){U}(\vec{x},t)$, where $\hat{U}(\vec{x},t)$ is the external potential in which the particle (fluid) moves (assumed to be non-differential for this article), the total energy density is then, 
\begin{equation}
    e = k_a +  q + u = k_c + q_r + u,
\end{equation}
and the integral over this density is the energy expectation value, which is constant if $U$ is time-independent.

\section{Local Conservation of Fluid Density and Energy Density}
\subsection{Time Evolution}
Now that we have this new type of potential energy in play, it will be helpful to consider how it moves within the fluid, and how energy is transferred from one type to another.  To begin, we return to the Schr\"{o}dinger equation, and left-multiply by $\psi^*$ to get the evolution equations for the densities,
\begin{equation}
    i\hbar \psi^* \frac{\partial \psi}{\partial t} = -\frac{\hbar^2}{2m}\psi^*\hat{K}\psi + \psi^*\hat{U}\psi \label{Sden}
\end{equation}
\begin{equation}
   =- \frac{\partial S}{\partial t}R^2 + i\hbar R \frac{\partial R}{\partial t}     \nonumber
\end{equation}
\begin{equation}
    = -\frac{\hbar^2}{2m}R     \nabla^2 R +\frac{1}{2m}R^2(\vec{\nabla}S \cdot \vec{\nabla}S)+ R^2U -  \frac{i\hbar}{2m}\big(2\vec{\nabla}R\cdot\vec{\nabla} S   +  R\nabla^2S\big)    \nonumber
\end{equation}

As discussed, the imaginary part gives the continuity equation for the fluid density, which can be seen from,
\begin{equation}
    \frac{\partial \rho}{\partial t} = - \vec{\nabla}\cdot\vec{j} = - \frac{1}{m}\vec{\nabla}\cdot(\rho \vec{\nabla}S) = - \frac{1}{m}\big( \vec{\nabla}\rho\cdot \vec{\nabla}S  + \rho\nabla^2 S \big)
\end{equation}
\begin{equation}
     = - \frac{1}{m}\big( 2R\vec{\nabla}R\cdot \vec{\nabla}S  + R^2\nabla^2 S\big) = 2R \frac{\partial R}{\partial t} , \nonumber
\end{equation}
from which it follows that,
\begin{equation}
    \frac{\partial R}{\partial t} = -\frac{1}{2m}\big( 2\vec{\nabla}R\cdot \vec{\nabla}S  + R\nabla^2 S\big) = - \frac{\vec{\nabla}\cdot(R^2 \vec{\nabla}S)}{2mR}, \label{RT}
\end{equation}
so we have the evolution equation for $R$, and the amount of fluid is a locally conserved quantity.

The real part of Eq. \ref{Sden} then gives us the evolution equation for $S$,
\begin{equation}
    \frac{\partial S}{\partial t} =  -\Big( \frac{1}{2m}(\vec{\nabla}S \cdot \vec{\nabla}S)   -\frac{\hbar^2}{2m}\frac{\nabla^2 R}{R} + U\Big),
\end{equation}
from which we can get the evolution for $\vec{\nabla}S$, which will be useful when we consider the time evolution of the energy densities,
\begin{equation}
    \frac{\partial \vec{\nabla}S}{\partial t} =  -\vec{\nabla} \bigg( \frac{1}{2m}(\vec{\nabla}S \cdot \vec{\nabla}S)   -\frac{\hbar^2}{2m}\frac{\nabla^2R}{R} + U\bigg). \label{ST}
\end{equation}

\subsection{Energy Density Continuity Equations}
Now, the continuity equations for the kinetic energy density, quantum potential energy density, and external potential energy density are, respectively,
\begin{equation}
\begin{array}{ccccc}
       \frac{\partial k_a}{\partial t} + \vec{\nabla}\cdot \vec{j}_k = \kappa, & & \frac{\partial q}{\partial t} + \vec{\nabla}\cdot \vec{j}_q = \alpha,  &  &    \frac{\partial u}{\partial t} + \vec{\nabla}\cdot \vec{j}_u = \beta.
\end{array} 
\end{equation}
where $\kappa$, $\alpha$, and $\beta$ represents the source/sink terms, and $\vec{j}_k \equiv k_a\frac{\vec{\nabla}S}{m} $ , $\vec{j}_u \equiv u\frac{\vec{\nabla}S}{m} $, and $\vec{j}_q \equiv q\Big(\frac{\vec{\nabla}S }{m} + \vec{v} \Big)$ are the energy current densities, where the kinetic and potential energy are external properties of the particles in the fluid, and thus their densities flow with the average fluid velocity $\frac{\vec{\nabla}S}{m}$.  The (internal) quantum potential energy density flows with the fluid, but can also flow through the fluid (from particle to particle) with an additional current $q\vec{v}$, which we have not yet defined.  

Note that we have begun with the case that all of the quantum potential energy $q$ is internal, and can flow between particles in the fluid, which is inconsistent with our model, but this still lays the mathematical groundwork for the case of our model, where only the reduced quantum potential energy $q_r$ is internal and flows between fluid particles with relative current $q_r\vec{v}_r$.

We will show that in either case, the requirement that energy is locally conserved fixes this additional current up to an additional divergence-free term.  This is consistent with prior analysis showing that, without some correction, the energy is not locally conserved in general \cite{holland1995quantum}.

So starting with the $q\vec{v}$ case, the three source/sink terms are, 
\begin{equation}
    \kappa = \frac{1}{m}R^2\vec{\nabla}S\cdot\frac{\partial \vec{\nabla}S}{\partial t} + \frac{1}{m^2}R^2 |\vec{\nabla}S|\vec{\nabla}S \cdot \vec{\nabla}|\vec{\nabla}S|
\end{equation}
\begin{equation}
     = \textcolor{purple}{\frac{\hbar^2}{2m} R^2\vec{\nabla}S \cdot   \vec{\nabla}\Big(\frac{\nabla^2R}{mR}\Big)}  - \textcolor{blue}{\frac{1}{m}R^2\vec{\nabla}S\cdot\vec{\nabla}U  }  ,   \nonumber
\end{equation}
\begin{equation}
    \alpha = -\frac{\hbar^2}{2m}\bigg(R\nabla^2\frac{\partial R}{\partial t} - \frac{\partial R}{\partial t}\nabla^2R +\textcolor{purple}{R^2\vec{\nabla}S \cdot   \vec{\nabla}\Big(\frac{\nabla^2R}{mR}\Big)} + \vec{\nabla}\cdot\big(\vec{v}R\nabla^2R\big) \bigg) ,
\end{equation}

and

\begin{equation}
    \beta = R^2\frac{\partial U}{\partial t} + \textcolor{blue}{\frac{1}{m}R^2\vec{\nabla}S \cdot \vec{\nabla}U} ,
\end{equation}
where we have used the continuity equation $\frac{\partial R^2}{\partial t} + \vec{\nabla}\cdot\Big( \frac{R^2\vec{\nabla}S}{m} \Big)=0$ to eliminate some terms in these expressions, and substituted in Eq. \ref{ST}.  The colors in the three equations above show the matched terms that represent local energy exchanges between the kinetic energy and the potential energies.

\subsection{Local Energy Conservation}

If we add all of the source/sink terms up, we see that the matched terms cancel as expected, but a few terms remain,
\begin{equation}
    \kappa + \alpha + \beta =  -\frac{\hbar^2}{2m}\bigg(R\nabla^2\frac{\partial R}{\partial t} - \frac{\partial R}{\partial t}\nabla^2R  + \vec{\nabla}\cdot\big(\vec{v}R\nabla^2R  \big)  \bigg)  +  R^2\frac{\partial U}{\partial t}. \label{localsources}
\end{equation}

If energy is to be locally conserved when $\frac{\partial U}{\partial t}=0$, then the term in parentheses must be zero.  Substituting in Eq. \ref{RT} we have,
\begin{equation}
 0 = -R\nabla^2 \bigg(\frac{\vec{\nabla}\cdot(R^2 \vec{\nabla}S)}{2mR}\bigg) + \bigg(\frac{\vec{\nabla}\cdot(R^2 \vec{\nabla}S)}{2mR}\bigg)\nabla^2R + \vec{\nabla}\cdot\big(\vec{v}R\nabla^2R  \big)
\end{equation}
\begin{equation}
    = \vec{\nabla}\cdot \bigg(- R \vec{\nabla}\bigg(\frac{\vec{\nabla}\cdot(R^2 \vec{\nabla}S)}{2mR}\bigg) + \vec{\nabla}R\bigg(\frac{\vec{\nabla}\cdot(R^2 \vec{\nabla}S)}{2mR}      \bigg) \bigg) + \vec{\nabla}\cdot\big(\vec{v}R\nabla^2R  \big), \nonumber
\end{equation}
\begin{equation}
    0 = \vec{\nabla}\cdot \bigg(- R^2  \vec{\nabla}\bigg(\frac{\vec{\nabla}\cdot(R^2 \vec{\nabla}S)}{2mR^2} \bigg)        + \vec{v}R\nabla^2R  \bigg) . \nonumber
\end{equation}
Thus we have the additional current,
\begin{equation}
    q\vec{v} = \frac{\hbar^2}{2m} R^2  \vec{\nabla}\bigg(\frac{\vec{\nabla}\cdot(R^2 \vec{\nabla}S)}{2mR^2} \bigg) +\vec{F},
\end{equation}
as the condition for local energy conservation in the quantum fluid, where $\vec{F}$ is any field satisfying $\vec{\nabla}\cdot \vec{F}=0$.  If we assume a current that satisfies this condition then source/sink term for the quantum potential energy density simplifies to,
\begin{equation}
    \textcolor{purple}{\alpha_\textrm{lec}} = \textcolor{purple}{-\frac{\hbar^2}{2m} R^2\vec{\nabla}S \cdot   \vec{\nabla}\Big(\frac{\nabla^2R}{mR}\Big)} 
\end{equation}

Now we move on to the case where only $q_r$ in an internal energy that can flow between particles in the fluid, while $k_s$ is now treated as external energy that only flows with the fluid.  We can use the calculation we've already done as a shortcut for finding the additional current $q_r\vec{v}_r$ need for local energy conservation.  The continuity equations for $k_s$ and $q_r$ are,
\begin{equation}
    \begin{array}{ccc}
      \frac{\partial k_s}{\partial t} + \vec{\nabla}\cdot\big(k_s\frac{\vec{\nabla}S}{m}\big) = \alpha_s,   &  & \frac{\partial q_r}{\partial t} + \vec{\nabla}\cdot\Big(q_r\big(\frac{\vec{\nabla}S}{m}  +\vec{v}\big)\Big) = \alpha_r   ,
    \end{array}
\end{equation}
where $\alpha_s$ and $\alpha_r$ are the source/sink terms, and $\vec{v}$ is again treated as unknown.  Now, noting that $\alpha_0 \equiv \alpha_s + \alpha_r = \alpha - \vec{\nabla}\cdot\big(k_s\vec{v}\big)$, where this is the old $\alpha$ from Eq. \ref{localsources}, the new sum of sinks and sources is,
\begin{equation}
    \kappa + \alpha_0 + \beta = \kappa + \alpha + \beta - \vec{\nabla}\cdot\big(k_s\vec{v}\big) 
\end{equation}
\begin{equation}
    = -\frac{\hbar^2}{2m}\bigg(R\nabla^2\frac{\partial R}{\partial t} - \frac{\partial R}{\partial t}\nabla^2R  + \vec{\nabla}\cdot\big(\vec{v}(R\nabla^2R   + \vec{\nabla}R\cdot\vec{\nabla}R)\big)  \bigg)  +  R^2\frac{\partial U}{\partial t}.\nonumber
\end{equation}
Following the same reasoning as before, to locally conserve energy we must now have,
\begin{equation}
    0 = \vec{\nabla}\cdot \bigg(- R^2  \vec{\nabla}\bigg(\frac{\vec{\nabla}\cdot(R^2 \vec{\nabla}S)}{2mR^2} \bigg)        + \vec{v}\big(\vec{\nabla}\cdot (R\vec{\nabla}R)\big)  \bigg), 
\end{equation}
and noting $\nabla^2(R^2) =  2\big(\vec{\nabla}\cdot (R\vec{\nabla}R)\big)$, the new solutions are,
\begin{equation}
    q_r\vec{v}_r = \frac{\hbar^2}{2m}R^2  \vec{\nabla}\bigg(\frac{\vec{\nabla}\cdot(R^2 \vec{\nabla}S)}{mR^2} \bigg) + \vec{F},
\end{equation}
for any field $\vec{F}$ such that $\vec{\nabla}\cdot\vec{F} = 0$.

\subsection{Energy Exchanges}
From $\kappa$, $\alpha$, and $\beta$ we can see the power density for local transfers between the kinetic, quantum potential, and external potential energies.

The term,
\begin{equation}
    p_u = R^2 \frac{\partial U}{\partial t},
\end{equation}
represents the power delivered to the fluid by a time-dependent external potential.
Next, we can identify the term,
\begin{equation}
    \textcolor{blue}{p_{k_au}} = -\textcolor{blue}{R^2\vec{\nabla} S\cdot  \vec{\nabla}U},
\end{equation}
which is the power density being converted from external potential to kinetic energy.
Lastly we have,
\begin{equation}
    \textcolor{purple}{p_{k_aq}} = \textcolor{purple}{\frac{\hbar^2}{2m^2} R^2\vec{\nabla}S \cdot   \vec{\nabla}\Big(\frac{\nabla^2R}{R}\Big)},
\end{equation}
which is the power density being converted from quantum potential to kinetic energy.

Now, to separate the quantum potential into the symmetric kinetic energy and reduced quantum potential, and identify how energy flows between them, and how it flows out to the average kinetic energy.  Assuming the local energy conservation condition is satisfied, we have to $\alpha_r = \alpha|_{\vec{v} = \vec{v}_r} - \textcolor{teal}{\alpha_s} - \vec{\nabla}\cdot(k_s\vec{v}_r)$, so we can see that $\alpha_r$ has the terms from $\alpha$ plus a direct energy exchange $\textcolor{teal}{\alpha_s}$ with the symmetric kinetic energy, where the power density delivered from the reduced quantum potential density to the symmetric kinetic energy density is,
\begin{equation}
   \textcolor{teal}{p_{q_rk_s}} = \textcolor{teal}{\alpha_s}  = \textcolor{teal}{\frac{\hbar^2}{2m^2}\bigg( \vec{\nabla}\cdot\Big(\vec{\nabla}S(\vec{\nabla}R\cdot\vec{\nabla}R)\Big)    -\vec{\nabla}R\cdot \vec{\nabla}\Big(\frac{\vec{\nabla}\cdot(R^2\vec{\nabla}S)}{R}\Big)  \bigg)}.
\end{equation}
This gives us
\begin{equation}
    \alpha_{r,\textrm{lec}} = -\textcolor{purple}{\frac{\hbar^2}{2m}R^2\vec{\nabla}S \cdot   \vec{\nabla}\Big(\frac{\nabla^2R}{mR}\Big)} 
\end{equation}
\begin{equation}
    -  \textcolor{teal}{\frac{\hbar^2}{2m^2}\bigg( \vec{\nabla}\cdot\Big(\vec{\nabla}S(\vec{\nabla}R\cdot\vec{\nabla}R)\Big)    -\vec{\nabla}R\cdot \vec{\nabla}\Big(\frac{\vec{\nabla}\cdot(R^2\vec{\nabla}S)}{R}\Big)  \bigg)}, \nonumber
\end{equation}
and thus the average kinetic energy is coupled to the reduced quantum potential, but not the symmetric kinetic energy, and the power density delivered from the reduced quantum potential to the average kinetic energy is,
\begin{equation}
    \textcolor{purple}{p_{k_aq_r}} =\textcolor{purple}{p_{k_aq}}= \textcolor{purple}{\frac{\hbar^2}{2m^2} R^2\vec{\nabla}S \cdot   \vec{\nabla}\Big(\frac{\nabla^2R}{R}\Big)},
\end{equation}
The full pattern of local energy exchanges is then,
\begin{equation}
    k_s \textcolor{teal}{\longleftrightarrow} q_r \textcolor{purple}{\longleftrightarrow} k_a \textcolor{blue}{\longleftrightarrow} u,
\end{equation}
and for our two groupings it reduces to,
\begin{equation}
    \begin{array}{ccc}
        q \textcolor{purple}{\longleftrightarrow}k_a \textcolor{blue}{\longleftrightarrow} u,    & \textrm{ or, }  & q_r \textcolor{black}{\longleftrightarrow}k_c \textcolor{blue}{\longleftrightarrow} u,
    \end{array}
\end{equation}
where $\longleftrightarrow$ includes both $\textcolor{teal}{\longleftrightarrow}$ and $\textcolor{purple}{\longleftrightarrow}$.

\section{What does the quantum potential do?}

Once we understand the quantum potential as a new type of local interaction within the fluid, we can start to look at how this interaction causes the fluid to behave in ways that a classical fluid never would.  Two important examples are how it explains phenomena like tunneling, where fluid is found in a classically forbidden region, and its role in quantum interference within the fluid.

The quantum potential has a tendency to cancel out jumps in the external potential, so for the fluid, it is as though the jumps aren't even there, or not entirely there.  To see this, consider the 1D Schr\"{o}dinger Equation with a constant external potential, $U(x,t) = U_0\Theta(x)$, where the $\Theta(x)$ is the Heaviside step function.  The boundary conditions at the finite discontinuity are $\psi_l(0,t) = \psi_r(0,t)$, from which it follows that $R_l(0,t)=R_r(0,t)$ and $e^{-iS_l(0,t)/\hbar}=e^{-iS_r(0,t)/\hbar}$, and $\vec{\nabla}\psi_l(0,t) = \vec{\nabla}\psi_r(0,t)$.  We begin from the Schr\"{o}dinger Equations for the left ($l$) and right ($r$) sides of the discontinuity,
\begin{equation}
    \begin{array}{c}
        \psi_l^*\sum_n E_n \psi_{n,l}e^{-iE_nt/\hbar} =  \psi_l^*\hat{K}\psi_l + U_lR_l^2, \\\\ \psi_r^*\sum_n E_n \psi_{n,r}e^{-iE_nt/\hbar} = \psi_r^*\hat{K}\psi_r + U_rR_r^2,
    \end{array}
\end{equation}
where $\psi(x,t) = \sum_n \psi_{n}(x)e^{-iE_nt/\hbar}$, and $\psi_n(x)$ are the energy eigenstates.  Next we consider the difference of these two equations, making use of the fact that $U_l=0$ and $U_r=U_0$, and substituting in the quantum kinetic energy density,
\begin{equation}
    \sum_n E_ne^{-iE_nt/\hbar} \big(\psi_r^*\psi_{n,r} - \psi_l^*\psi_{n,l} \big) = q_{r,r}-q_{r,l} + k_{a,r} - k_{a,l} + k_{s,r} - k_{s,l} + U_0R_r^2.  \label{Dis} 
\end{equation}
The boundary conditions apply to both the state and the eigenstates, so we can see that the term in the parentheses in Eq. \ref{Dis} is zero at the boundary.  Furthermore, recalling that $k_a = \frac{1}{2m}R^2|\vec{\nabla}S|^2$ and $k_s = \frac{\hbar^2}{2m}|\vec{\nabla}R|^2$, and using the identities,
\begin{equation}
    \vec{\nabla}R = \frac{1}{2}\Big(e^{-iS/\hbar}\vec{\nabla}\psi + e^{iS/\hbar}\vec{\nabla}\psi^*\Big) = \textrm{Re}\Big(e^{-iS/\hbar}\vec{\nabla}\psi\Big),
\end{equation}
and,
\begin{equation}
    \vec{\nabla}S = \frac{1}{2iR}\Big(e^{-iS/\hbar}\vec{\nabla}\psi - e^{iS/\hbar}\vec{\nabla}\psi^*\Big) =\frac{1}{R}\textrm{Im}\Big(e^{-iS/\hbar}\vec{\nabla}\psi\Big),
\end{equation}
the boundary conditions give us $k_{a,r} = k_{a,l}$ and $k_{s,r} = k_{s,l}$, so Eq. \ref{Dis} reduces to,
\begin{equation}
\begin{array}{ccc}
      q_{r,r}-q_{r,l} = -U_0R^2, & \textrm{ or, } &     Q_{r,r}-Q_{r,l} = -U_0. 
\end{array}
\end{equation}
Thus, as claimed, the discontinuity in the reduced quantum potential is equal and opposite to the discontinuity in the external potential, and so the total potential energy density $q_r +u$ is a continuous function, and by extension, so is $q+u$.

This effect is particularly relevant for tunneling through a rectangular barrier, and a detailed example showing the roles of the different types of energy during such tunneling is given in the Appendix.  While truly discontinuous potentials are probably nonphysical, they are still an interesting limiting case.  Perhaps most importantly, we have also proved that if the external potential is continuous, then so is the reduced quantum potential.

The Appendix also contains a detailed example of reflection off an infinite barrier.

\section{Discussion}

So far we have ignored a conceptual issue that may be problematic for this model, which is that the velocities and energies of the particles generally contain singularities, as does the additional current $q_r\vec{v}_r$ of the reduced quantum potential energy.  Despite this, all of the energy densities we have considered, and all of the currents associated with velocity $\vec{\nabla}S/m$, are well-behaved finite functions, which has allowed for our analysis.  Furthermore, while the fluid current, average kinetic energy density, and quantum potential energy density are all zero where the fluid density is zero, the symmetric velocity is singular, so the symmetric current, momentum, and kinetic energy densities, and the reduced quantum potential energy density can all be finite where the fluid density has an isolated zero (although not where it is smoothly zero).  The fact that there can be current, momentum, and energy where there are no fluid particles seems like a conceptual problem, but since this can only happen at isolated zeroes in the density, the overall amount of current, momentum, and energy at these locations is of measure zero, maybe it is nothing to worry about.

In general, the infinite velocities undermine the premise that this model is local, but we think this is likely to be an artifact of the nonrelativistic treatment we have used.  Our hope is that in a proper relativistic treatment, these singularities will vanish and the particle velocity in the fluid will be bounded by $c$ (this seems to work out in \cite{poirier2012trajectory} and \cite{ruiz2021direct}).  Restricting $\vec{v}_s$ in this way could also make it so that all current, momentum, and energy densities are zero where the fluid density is zero.  The issue of quantized vortices surrounding nodes in wavefunctions of many types has been studied extensively \cite{takabayasi1954hydrodynamical,takabayasi1955vector, takabayasi1983vortex,takabayasi1983hydrodynamical,hirschfelder1974quantized,hirschfelder1977angular,hirschfelder1974quantum,reddiger2023towards}, but this issue does not appear to be of particular importance for the analysis we have given here - although it is clearly important for understanding the detailed behavior of the fluid near the nodes.  These vortices and their mathematical underpinnings are also relevant to the treatment of vector potentials, the Aharonov-Bohm effect, and spin, which goes beyond the scope of this article. 

In conclusion, while the present model has many satisfying features, it also has some apparent problems that need to be addressed.  However, there is good reason to think that a relativistic picture will overcome these issues, since we are looking for a fluid model consistent with the local Heisenberg picture used in relativistic quantum field theory.  We also suspect that in the generalized treatment, the fluid particles will be of definite energy rather than definite mass, which will mix up several of the fluid properties we have considered here, and also allow for the treatment of massless particles.  In general, energy is a locally conserved quantity in (special) relativistic theories, so it is quite natural to think of it as a fluid anyway.  

In the local many worlds picture \cite{waegell2023local}, the conserved fluid of each quantum system is separated into multiple branches during an interaction between systems, each corresponding to a different outcome, and the relative amounts of fluid in each branch give rise to the Born rule probability of observing that outcome at macroscopic scale.  If each fluid particle is taken to be of definite energy (in a given frame), then this is just another manifestation of local conservation of energy, whereas in our present treatment it is effectively conservation of mass, which we should not expect to survive a generalization to special relativity.  The fluid particles should also possess definite values for any other locally conserved property, whether frame invariant like charge, or frame-dependent like momentum.

Finally, it is worth noting that many other works have explored different ways to interpret or modify Madelung's original fluid equations (or their Bohm equivalent) \cite{bohm1954model,schonberg1954hydrodynamical,spiegel1980fluid,salesi1996spin,delphenich2002geometric,pashaev2002resonance,caliari2004dissipation,poirier2010bohmian,poirier2004reconciling,nonnenmacher2005hamiltonian,wyatt2005quantum,trahan2006reconciling3,poirier2007reconciling,morato2007formation,poirier2008reconciling,poirier2008reconciling6,budiyono2009madelung,poirier2012action,poirier2012trajectory,heifetz2015toward,vadasz2016rendering,heifetz2016entropy,reddiger2017madelung,heim2018recursive,ruiz2021direct,finley2021refined,de2021making,finley2022fields,finley2022fluid,reddiger2023towards}, so even if the present model fails, there are plenty of alternative ideas available.  For an especially thorough review of the literature and discussion of some recurring issues, see \cite{reddiger2023towards}. \newline\newline
\textbf{Conflicts of Interest:}  There is no conflict of interest.\newline\newline
\textbf{Data Availability Statement:}  This is a theoretical work and the only numerical data of interest is given in the figures.

\printbibliography

\section{Appendix A: Tunneling and Interference Examples}

The simplest tunneling example is an infinite square well of width $2L$, and with a finite rectangular barrier of width $2a$ at its center (for other examples and simulations, see \cite{hirschfelder1974quantum,dewdney1982quantum,bittner2000quantum,trahan2006reconciling}).  If the two lowest energy eigenvalues, $E_1$ and $E_2$, are below the barrier height $U_0$, then the eigenstates are of the form,
\[ \psi_1(x) = \left\{  \begin{array}{lll}
A\sin(k_1(x+L)) && -L \leq x \leq -a \\
B\cosh(\kappa_1 x) && -a\leq x\leq a \\ 
A\sin(k_1 (L-x)) &&  a \leq x\leq L
\end{array}\right.,\]
and 
\[ \psi_2(x) = \left\{  \begin{array}{lll}
C\sin(k_2(x+L)) && -L \leq x \leq -a \\
D\sinh(\kappa_2 x) && -a\leq x\leq a \\ 
-C\sin(k_2 (L-x)) && a \leq x\leq L
\end{array}\right.,\]
where $A,B,C,$ and $D$ are fixed by the boundary conditions and normalization, and $k_n = \frac{\sqrt{2mE_n}}{\hbar}$, and $\kappa_n = \frac{\sqrt{2m(U_0 - E_n)}}{\hbar}$.  

If we prepare an initial state $\psi(x,0) = \frac{1}{\sqrt{2}}\big(\psi_1(x) + \psi_2(x)\big)$, most of the fluid is on one side of the barrier, and as time evolves it tunnels until most of it is on the other side, and then back.

\begin{figure}
    \centering
    \begin{tabular}{cc}
 \includegraphics[width=2.5in]{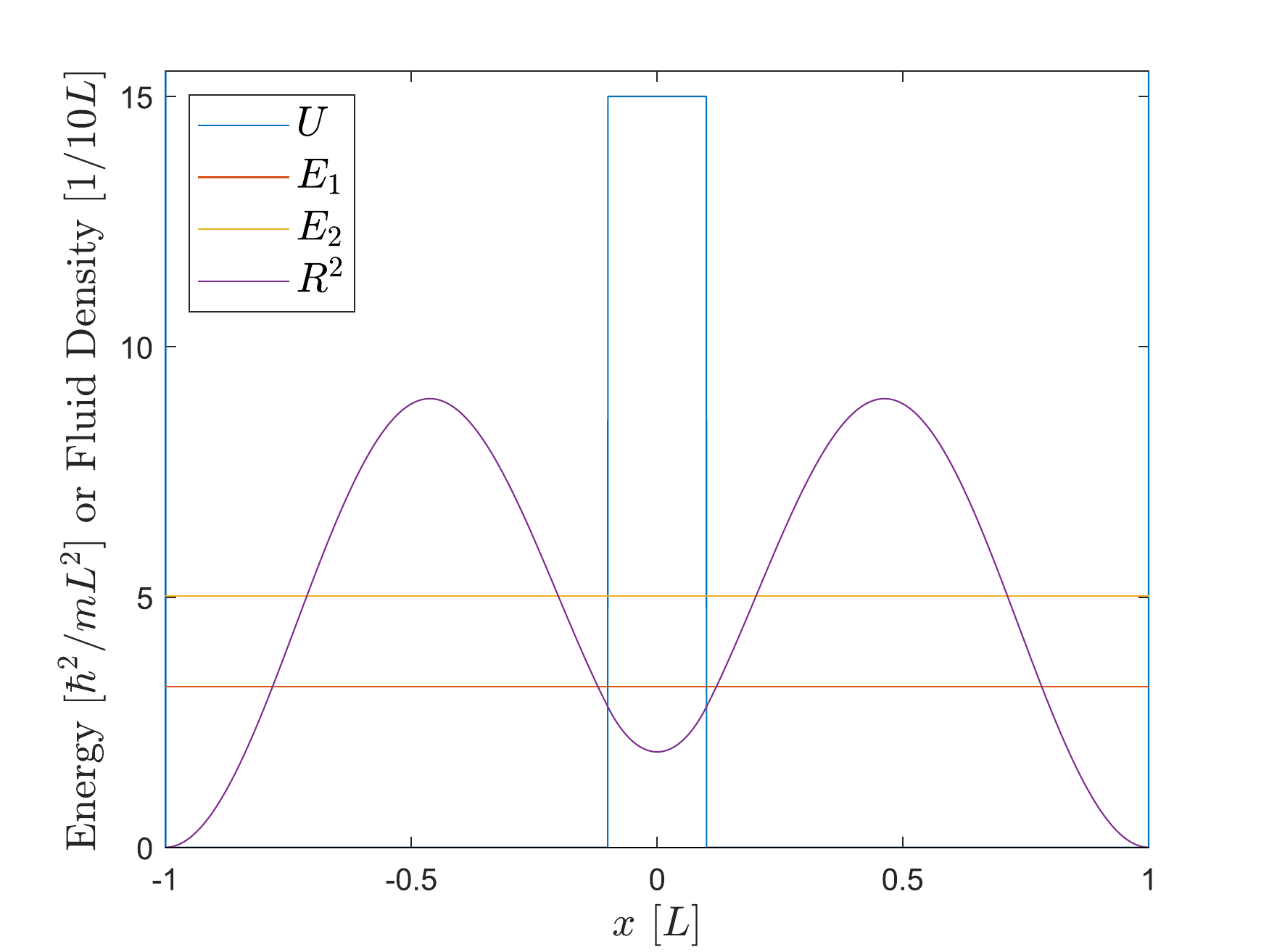} &\includegraphics[width=2.5in]{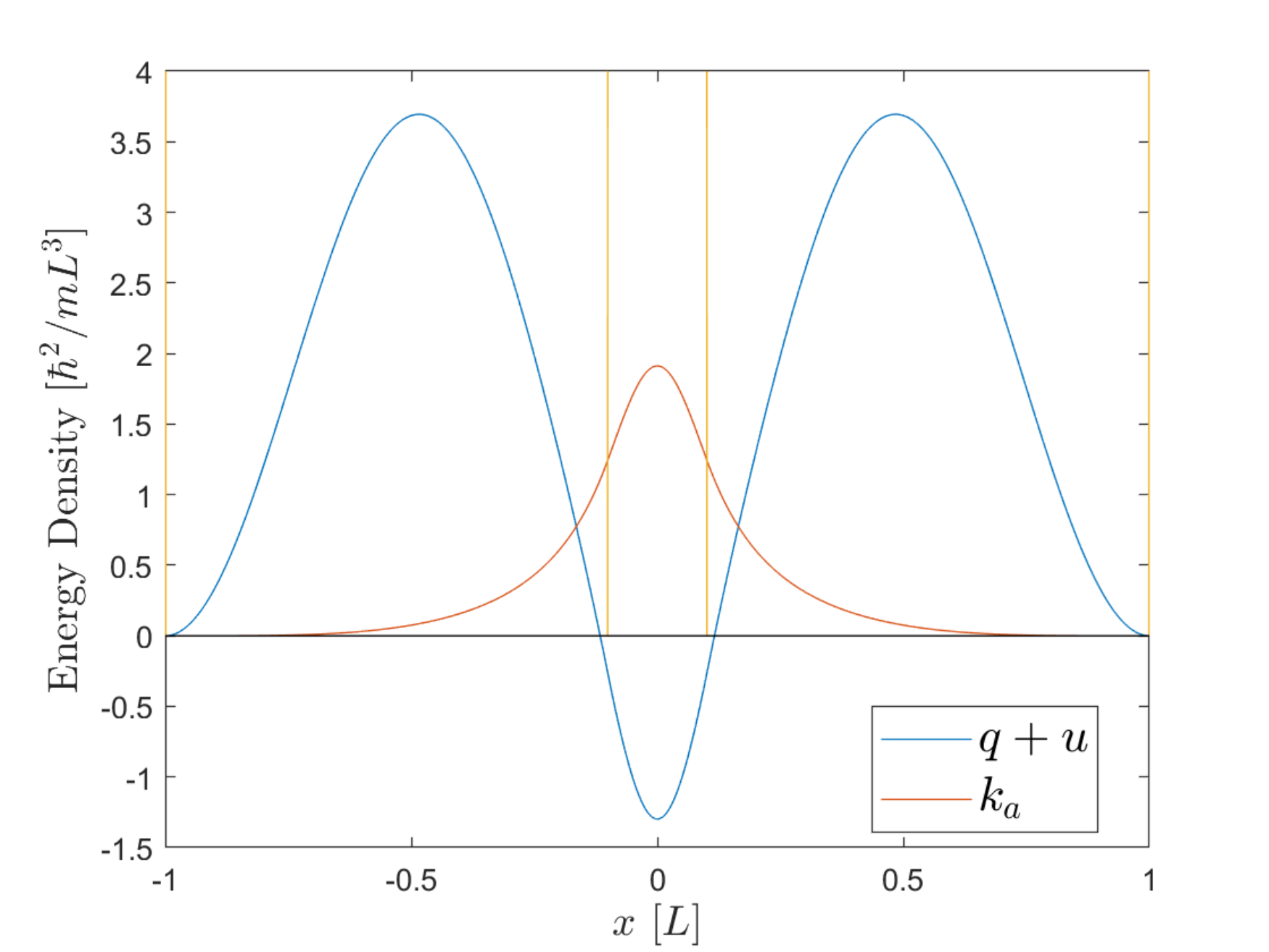}   \\
    
\includegraphics[width=2.5in]{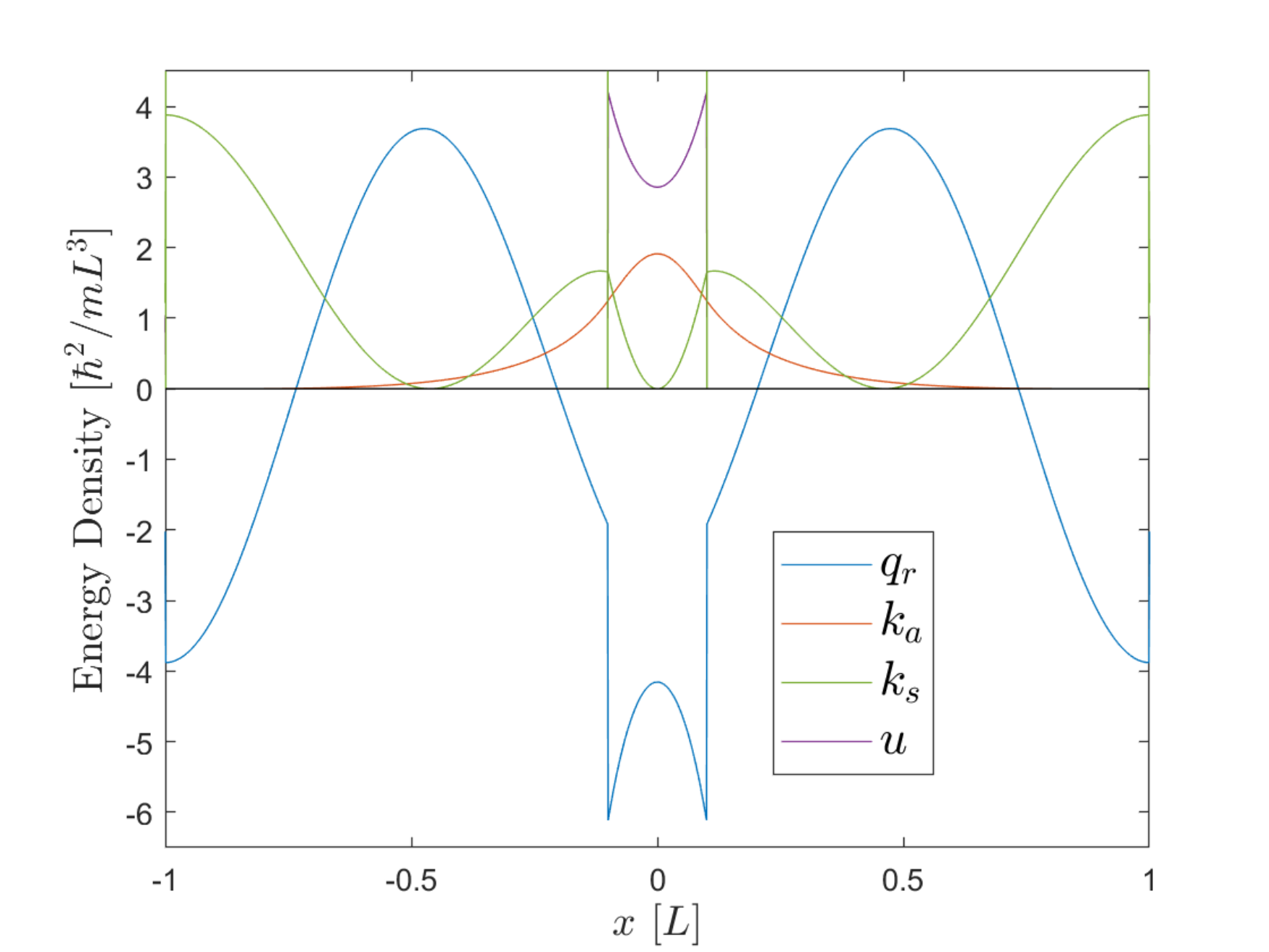} &\includegraphics[width=2.5in]{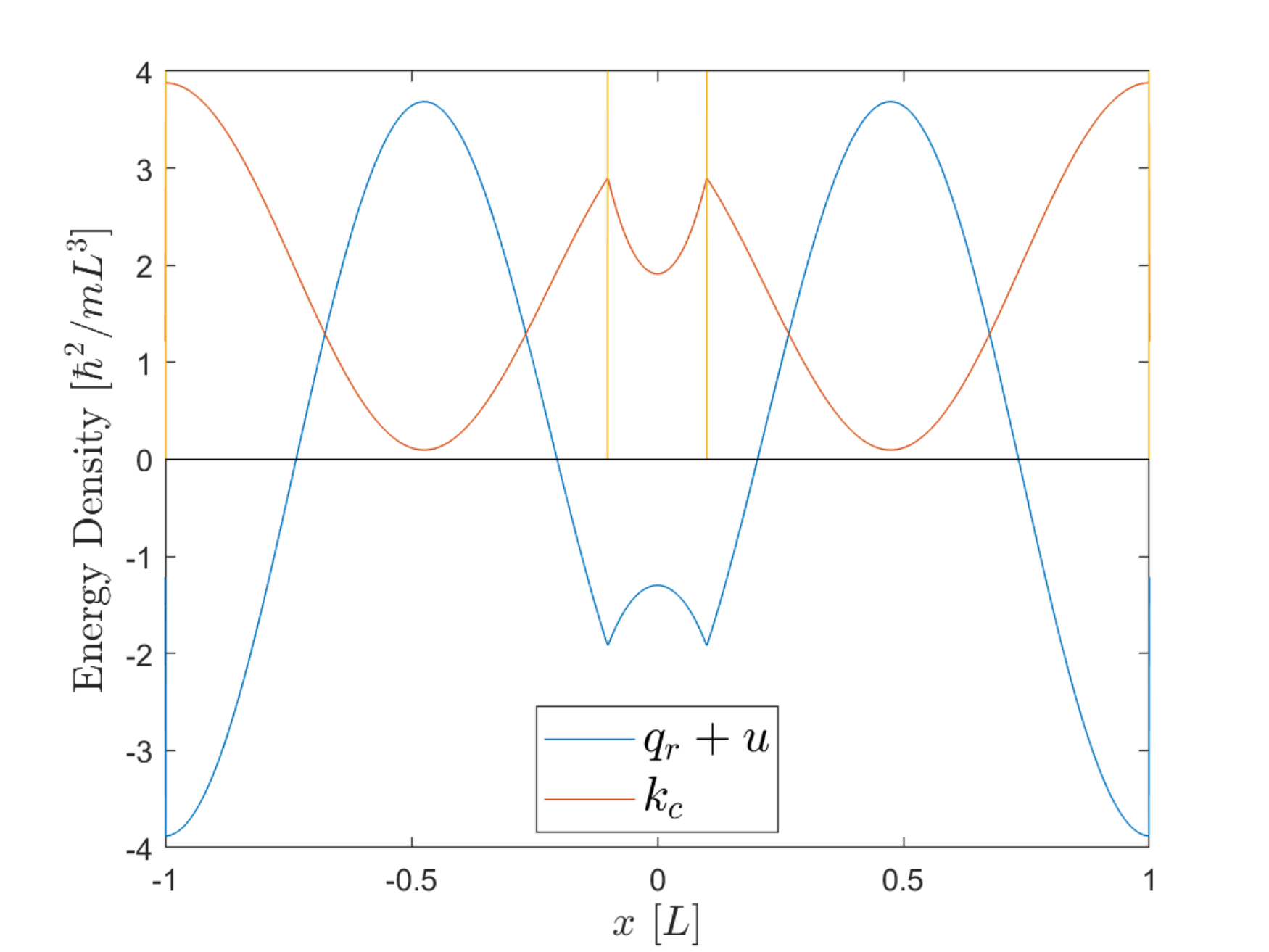}
     
\end{tabular}
    \caption{\small{\textbf{(top,left)} A double-well potential $U$ with infinite walls and a finite rectangular barrier in the center, the two lowest energy eigenvalues, $E_1$ and $E_2$, and eigenstates, $\psi_1(x)$ and $\psi_2(x)$, and the fluid density $R^2$ for the state $\psi(x,T/4) = (\psi_1(x)e^{-iE_1T/4\hbar} + \psi_2(x)e^{-iE_2T/4\hbar})/\sqrt{2}$, with period $T = 2\pi\hbar/(E_2-E_1)$.    
    \textbf{(bottom,left)} The reduced quantum potential, $q_r = -(\hbar^2/4m)\nabla^2 R^2$, average kinetic, $k_a = (1/2m)|\vec{\nabla}S|^2$, symmetric kinetic, $k_s =(\hbar^2/2m)|\vec{\nabla}R|^2$, and external potential, $u = R^2U$, energy densities, for the state $\psi(x,T/4)$.  Note the equal and opposite discontinuities in $q_r$ and $u$.
    \textbf{(top,right)} The standard grouping, with quantum potential energy density $q = q_r + k_s$ added to $u$ to form a single potential energy density, and the average kinetic energy density $k_a$ corresponding to the flow.  This picture seems to track $R^2$ and the current $R^2\vec{v}_a$ most intuitively, and shows increased kinetic energy where the fluid falls into a dip in the potential energy as it tunnels through the (cancelled) barrier.
    \textbf{(bottom,right)} The new grouping where the reduced quantum potential $q_r$ is added to $u$ to form a single potential energy density, and the average and symmetric kinetic energy densities are combined to form a single classical kinetic energy density, $k_c = k_a + k_s$.  This may be the more physically correct picture, where there is motion and kinetic energy within stationary states.  
    }}
    \label{TunDen}
\end{figure}

\begin{figure}
    \centering
    \includegraphics[width=5.5in]{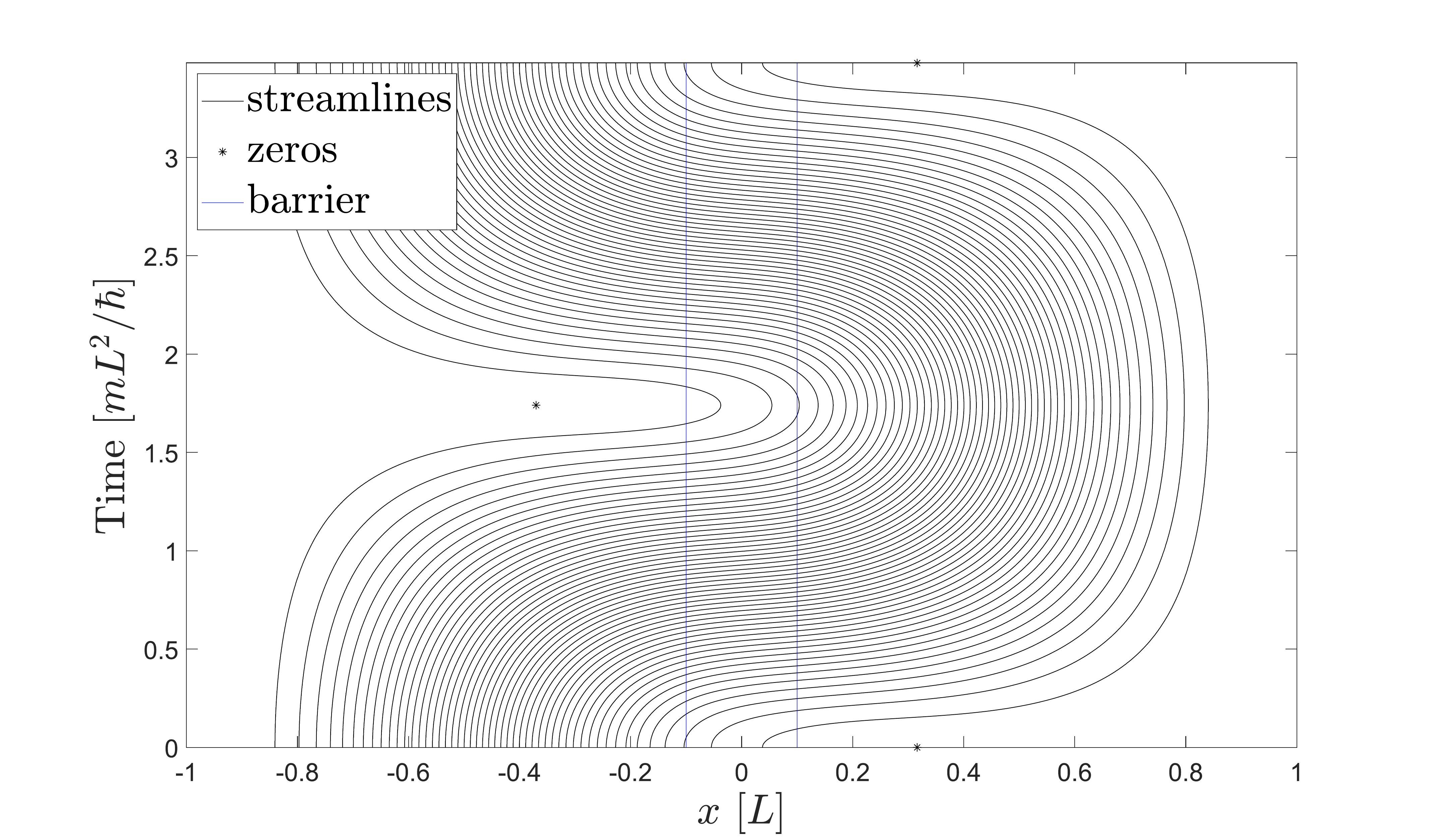}
    \caption{Fluid streamlines for one full period $T = 2\pi\hbar/(E_2-E_1)$ of the state $\psi(x,t) = (\psi_1(x)e^{-iE_1t/\hbar} + \psi_2(x)e^{-iE_2t/\hbar})/\sqrt{2}$.  The fluid density is proportional to the horizontal density of the streamlines.  The momentary zeros that form due to the interference of the two eigenstates are also shown.  An animation of this motion is given in the Supplementary Material, which shows particles that track the streamlines, the average kinetic energy density, $k_a$, and the potential energy density, $q + u$.  In the animation, it is apparent that the kinetic energy bump where the fluid falls into a dip in the potential energy at one zero in the fluid density, grows and moves through the barrier, and then shrinks and disappears at the zero that forms on the opposite side.  The kinetic bump and potential dip features seem to appear for dark interference fringes involving multiple energies, which makes their apparent connection to tunneling interesting.}
    \label{TunStr}
\end{figure}

The example in Fig. \ref{TunDen} shows the energy eigenvalues, $E_1 = 3.22 \hbar^2/mL^2$ and $E_2 = 5.02 \hbar^2/mL^2$ (found numerically), for the shown double-well potential with barrier height $U_0 = 15 \hbar^2/mL^2$, the fluid density, the four energy density terms $q_r$, $k_a$, $k_s$, and $u$, and the two alternative groupings, all in the middle of this tunneling process.  In both groupings, not only has the (reduced) quantum potential canceled out the barrier, but it has also created a depression beneath it, which is compensated by a bump in the kinetic energy of the fluid within the barrier - so we have a nearly classical explanation of tunneling.  The animation in the Supplementary Material shows the energy densities for the entire cycle, with streamline markers that move at velocity $\vec{v}_a$ along the fluid streamlines.  A $t$ vs. $x$ plot of the streamlines is given in Fig. \ref{TunStr}.

\begin{figure}
    \centering
    \begin{tabular}{cc}
 \includegraphics[width=2.5in]{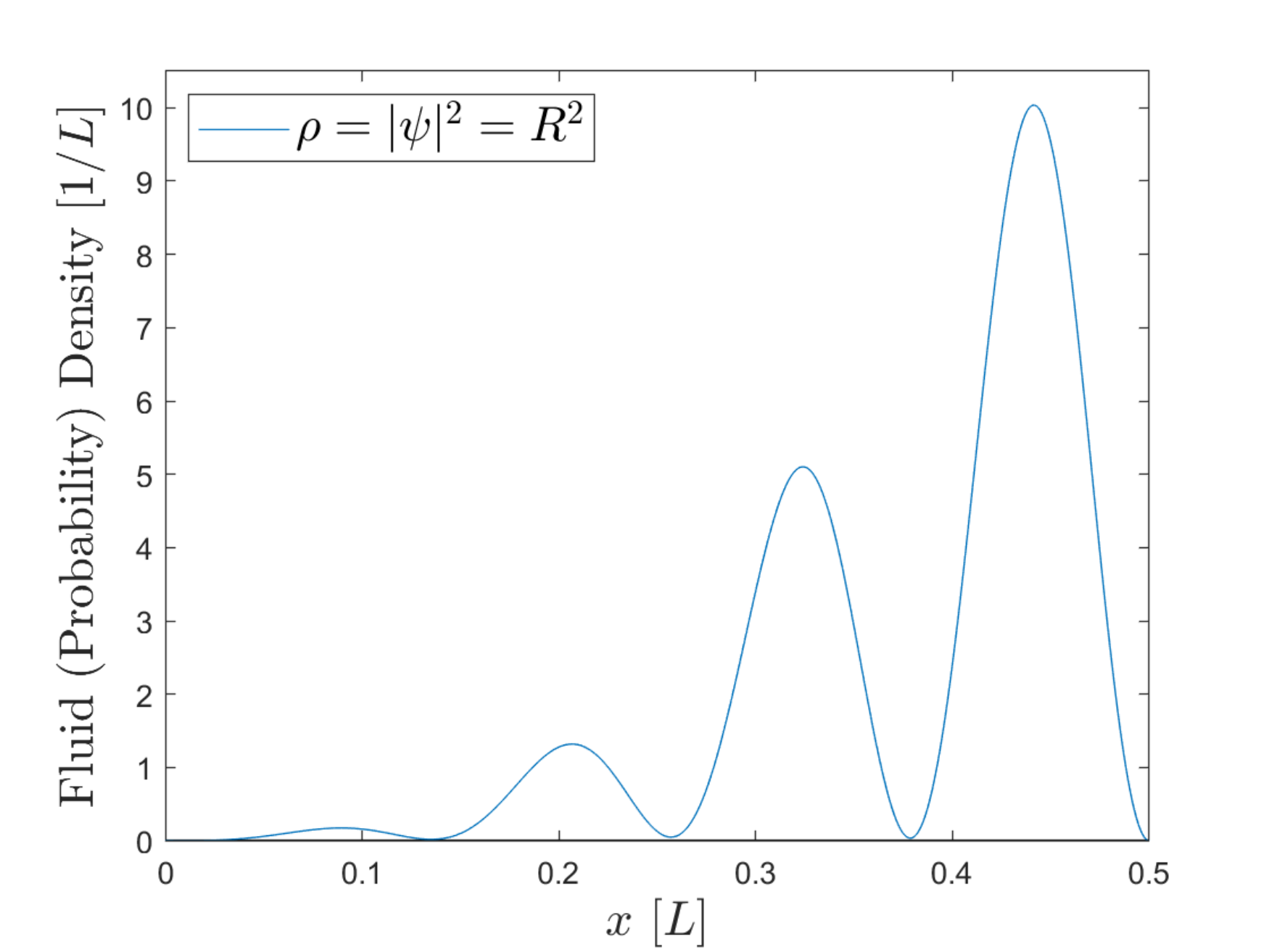} &\includegraphics[width=2.5in]{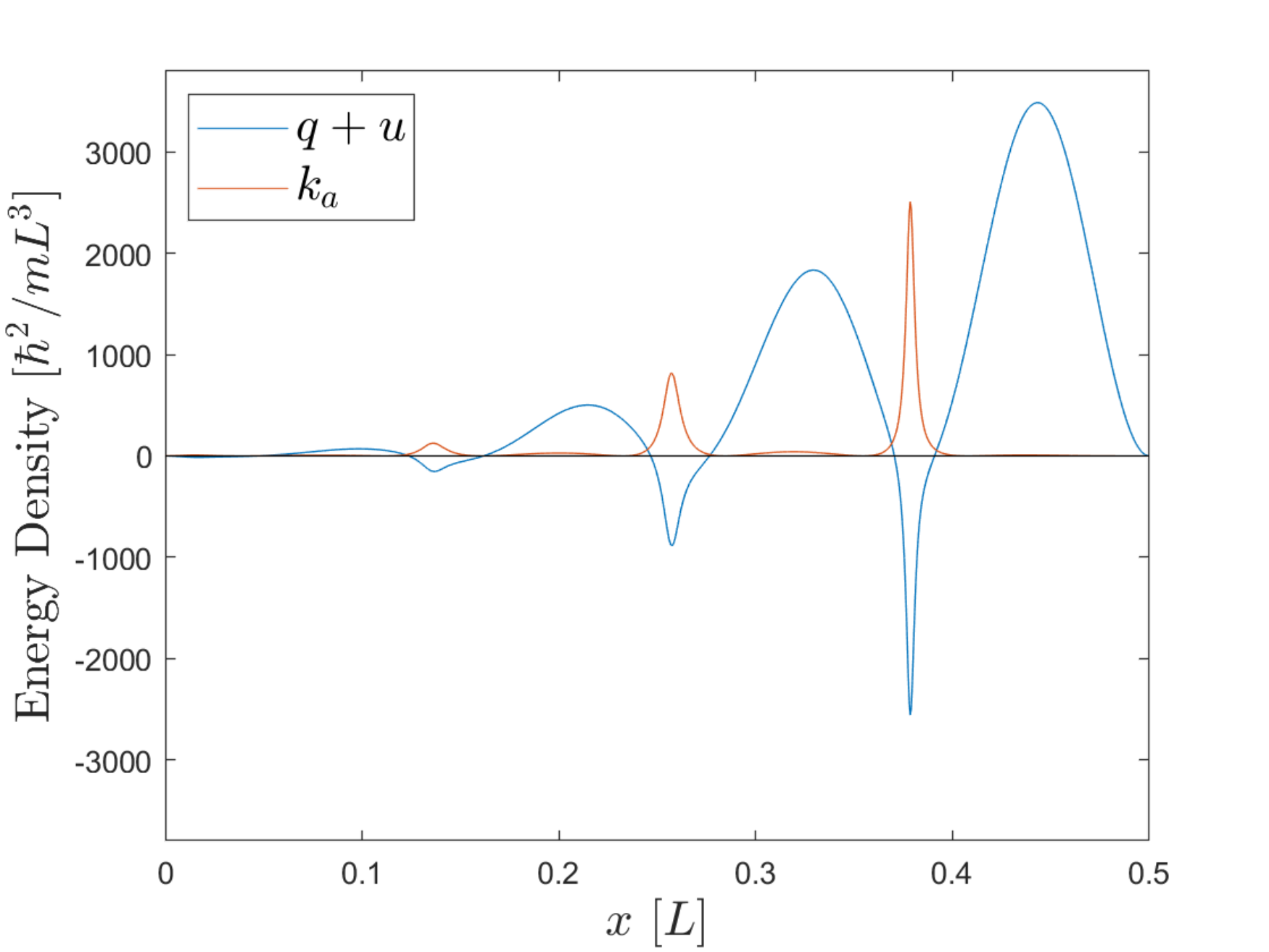}   \\
\includegraphics[width=2.5in]{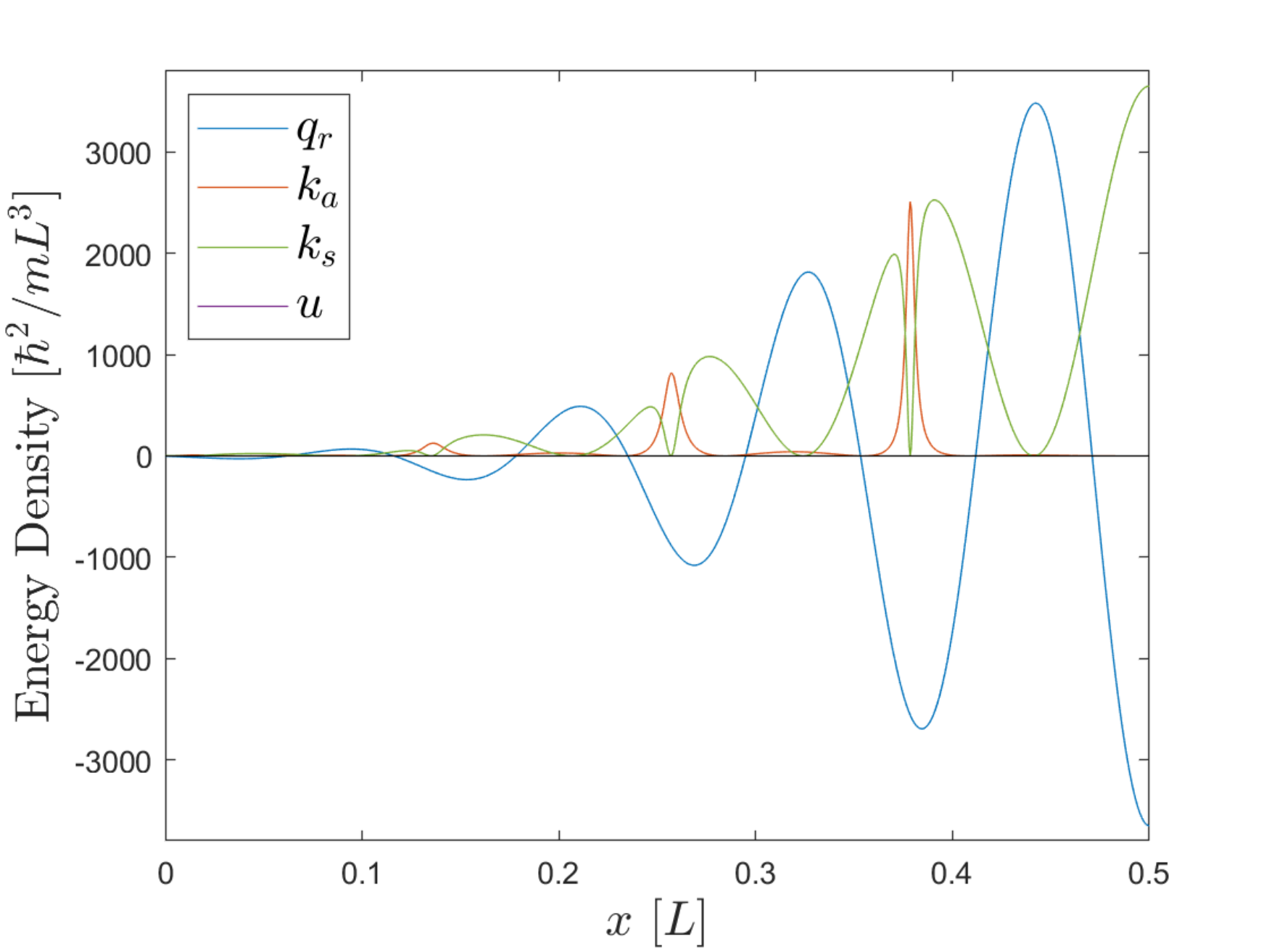} &\includegraphics[width=2.5in]{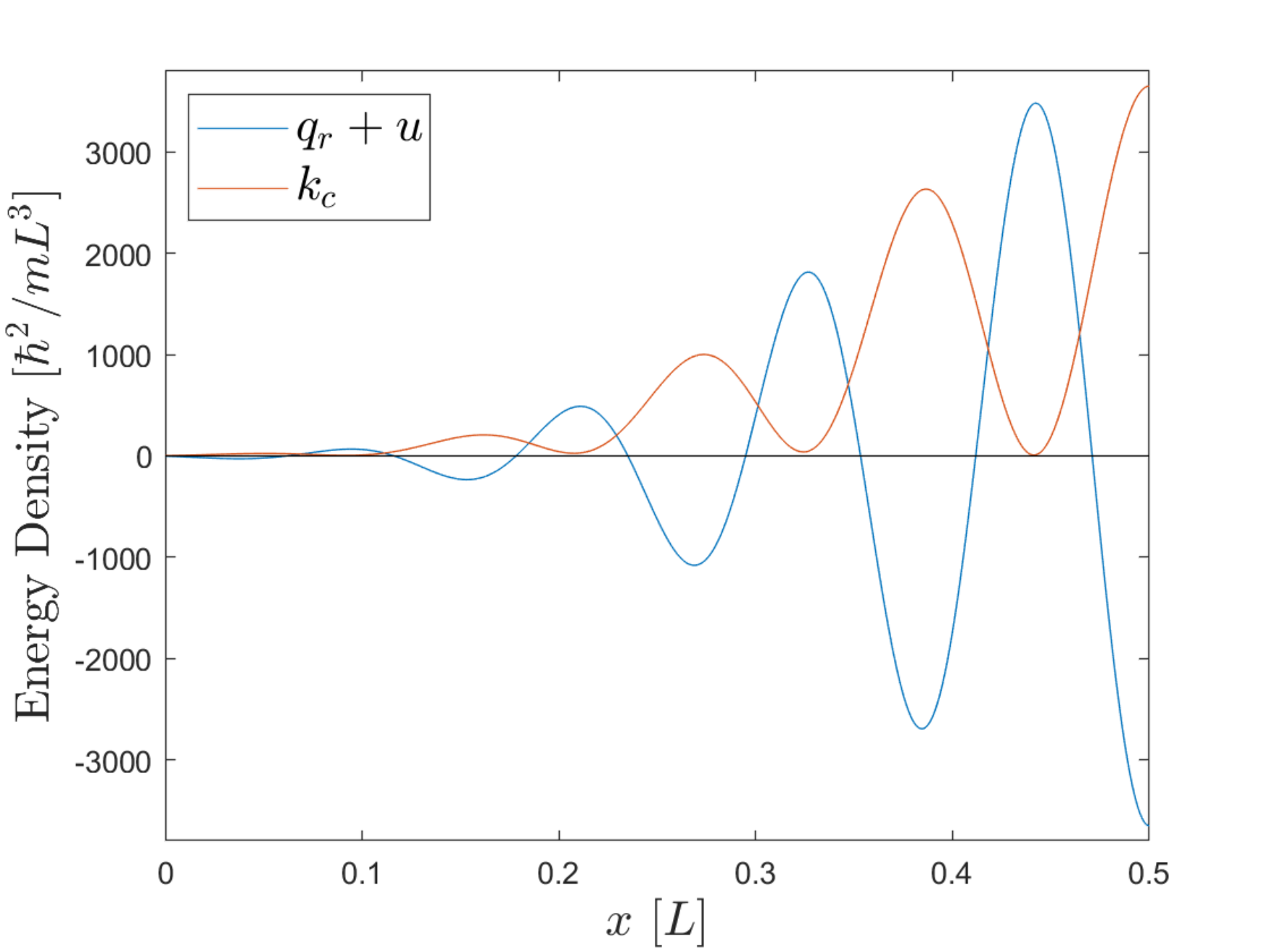}
     
\end{tabular}
    \caption{\small{\textbf{(top,left)} The fluid density $R^2$ at time $t_1=0.1907[mL^2/\hbar]$ for the Gaussian pulse prepared at $t=0$ as $\psi(x,0) = A e^{-x^2/(4\Delta x^2) + ip_0 x/\hbar}$, with $\Delta x =  L/(4\sqrt{10})$ and $p_0 = 25 \hbar/L$, and normalization constant $A$, as it reflects off an infinite barrier at $x=L/2$.  This moment has been chosen just before the dark interference fringes form, when the densities in the troughs are still slightly above zero.
    \textbf{(bottom,left)} The reduced quantum potential, $q_r = -(\hbar^2/4m)\nabla^2 R^2$, average kinetic, $k_a = (1/2m)|\vec{\nabla}S|^2$, symmetric kinetic, $k_s =(\hbar^2/2m)|\vec{\nabla}R|^2$, and external potential, $u = R^2U$, energy densities, for the state $\psi(x,t_1)$.  Note the equal and opposite spikes in $k_a$ and $k_s$.
    \textbf{(top,right)} The standard grouping, with quantum potential energy density $q = q_r + k_s$ added to $u$ to form a single potential energy density, and the average kinetic energy density $k_a$ corresponding to the flow.  This picture seems to track $R^2$ and the current $R^2\vec{v}_a$ most intuitively, and shows increased kinetic energy where the fluid falls into a dip in the potential energy as it approaches a dark interference fringe ($R^2=0$).  Interestingly, that extra average kinetic energy appears to truly come out of the symmetric kinetic energy in this case.
    \textbf{(bottom,right)} The new grouping where the reduced quantum potential $q_r$ is added to $u$ to form a single potential energy density, and the average and symmetric kinetic energy densities are combined to form a single classical kinetic energy density, $k_c = k_a + k_s$.  This may be the more physically correct picture, where there is motion and kinetic energy within stationary states.  It is interesting that in this case, the spikes cancel and we get a smoothly oscillating $k_c$ and $q_r$.
    }}
    \label{RefDen}
\end{figure}

\begin{figure}
    \centering
   \includegraphics[width=7in]{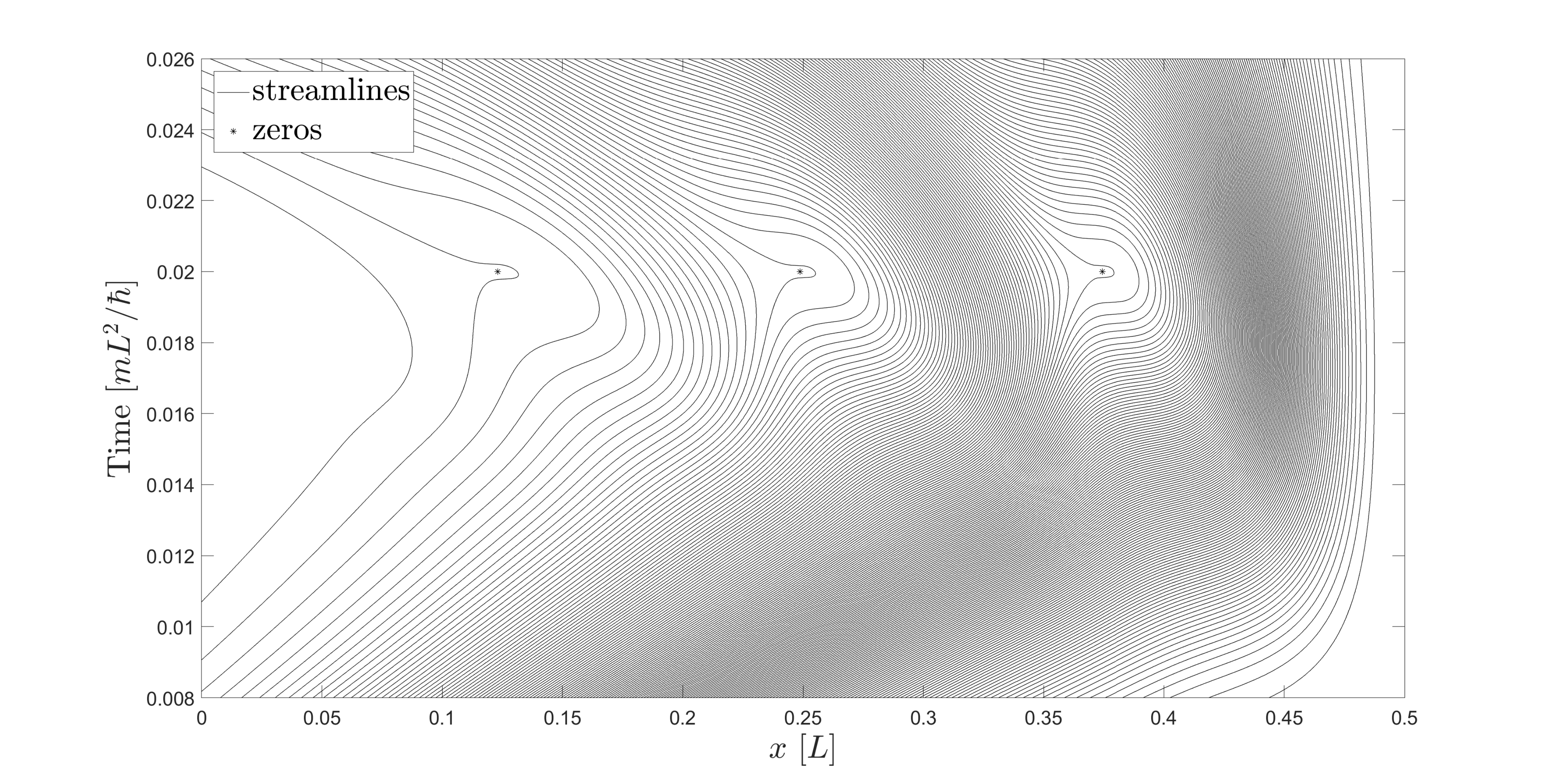}
    \caption{Fluid streamlines for a Gaussian pulse prepared at $t=0$ as $\psi(x,0) = A e^{-x^2/(4\Delta x^2) + ip_0 x/\hbar}$, with $\Delta x =  L/(4\sqrt{10})$ and $p_0 = 25 \hbar/L$, and normalization constant $A$, as it reflects off an infinite barrier at $x=L/2$.  The fluid density is proportional to the horizontal density of the streamlines.  Interference between the incoming and reflected portions of the pulse causes zeros in the fluid density to momentarily form half-way through the reflection process.  An animation of this motion is given in the Supplementary Material, which shows particles that track the streamlines, the average kinetic energy density, $k_a$, and the potential energy density, $q + u$.  It is interesting to note that as the zero in the fluid density forms, relatively few fluid particles approach it, and those that do, fall into the potential dip and are sped across.  However, particles that have not made it past by the time the zero forms are reflected back.}
    \label{RefStr}
\end{figure}

Another important example is the role of the quantum potential in quantum interference phenomena.  Here we consider a 1D Gaussian pulse that is reflecting off an infinite barrier, creating interference between the incoming and reflected parts of the pulse.  As the destructive fringes form and $R^2$ approaches zero, the average kinetic energy density $k_a$ forms a spike above each local minimum of $R^2$, and the symmetric kinetic energy density $k_s$ forms a dip at these same locations.  These spikes and dips appear as the destructive interference fringes (zeros) in the fluid density begin to form, vanish only at the moment the fringes appear, and then reappear as the fringes vanish.

The example in Fig. \ref{RefDen} shows the fluid density, the four energy density terms $q_r$, $k_a$, $k_s$, and $u$, and the two alternative groupings, just before the dark fringes form, as the Gaussian pulse,
\begin{equation}
    \psi(x,0) = A \exp{\bigg(-\frac{x^2}{4\Delta x^2} + i\frac{p_0 x}{\hbar}\bigg)},
\end{equation}
reflects off an infinite barrier at $x=0.5L$ (with $A$ a normalization constant).  At $t=0$ the pulse is centered at the origin, with mean momentum $p_0 = 25 \hbar/L$, and spread $\Delta x = L/(4\sqrt{10})$, and these plots are at $t=0.01907 (mL^2/\hbar)$, when the spikes and dips can be seen.  The middle of the reflection occurs at $t = 0.0200 (mL^2/\hbar)$, and the dark fringes form zeros in the fluid density, which can be seen in Fig. \ref{RefStr} along with fluid streamlines during the reflection.

However, if we consider the classical kinetic energy density $k_c = k_a + k_s$, the spike and dip exactly cancel, leaving a smoothly varying oscillation through the entire pulse, so the two types of kinetic energy also seem to lock naturally together, similarly to $q$ and $u$ in the tunneling case.  In either grouping, the minima of the fluid density correspond to depressions in the (reduced) quantum potential which are compensated by bumps in the kinetic energy.

The animation in the Supplementary Material shows the energy densities for the reflection process, with streamline markers that move at velocity $\vec{v}_a$.

Looking at both animations, it is interesting to note that the dip beneath the barrier in the tunneling case actually begins at a zero in the fluid density outside the barrier and then grows and moves to the center, so there appears to be some connection between dark interference fringes (zeros) and the presence of fluid in classically forbidden regions.  In both cases, regions where fluid velocities exceed our classical expectations have corresponding dips in the quantum potential.

\section{Appendix B: From Configuration Space to 3-Space}

Madelung's original analysis applies to a single quantum particle, and interprets the probability density and current for that one quantum particle as the literal density and current of a continuum fluid in 3-space.  If we assume that this fluid description is the continuum limit of a classical fluid that is itself composed of many classical particles (called `fluid particles' throughout this article), each with a definite position, momentum, and energy, then we have an essentially infinite number of these fluid particles corresponding to a single quantum particle (e.g., one electron).  One way to interpret this is to consider the fluid particles of the classical fluid as a many-worlds representation of the single quantum particle, and this leads naturally to the Born rule, as discussed in the main text.

In classical fluid mechanics, the configuration space of $N$ fluid particles is $3N$-dimensional, but it is still understood that each fluid particle lives in 3-space, with a well defined position and velocity, and the continuum fluid equations likewise live in 3-space.  

Once we introduce two quantum particles (e.g., two electrons) which may be entangled, the probability density and current in standard quantum theory lives in a 6-dimensional configuration space, and there is no straightforward way to represent this in 3-space.  If we want to interpret the probability density and current for our two quantum particles as the literal density and current of a continuum fluid, that fluid now appears to live in 6-space, and the position and velocity of a fluid particle actually represents a pair of positions and a pair of velocities in 3-space.  Entanglement correlations between the two quantum particles appear to make it impossible to separate this into a story about a classical fluid for one quantum particle in 3-space, and another about a classical fluid for the other quantum particle in 3-space.

Despite the apparent problem, it has recently been shown \cite{waegell2023local} that it is possible to separate this into \textit{multiple} distinct stories about classical fluids in 3-space which collectively represent one quantum particle, and \textit{multiple} distinct stories about classical fluids in 3-space which collectively represent the other quantum particle.  This representation contains all of the same information about phases and entanglement correlation as the standard quantum theory, and leads to the same empirical predictions - although because the theory is local, we must be careful about when and where the empirical data can be observed.

For any number of quantum systems with a joint Hilbert space of dimension $D$, each individual quantum system will generally be represented by $D$ distinct classical fluids in 3-space, each indexed by an orthogonal direction in the Hilbert space.  This is simplest for the case of two spin-1/2 quantum particles which are not entangled with their spatial degrees of freedom, since then there are just four distinct fluids for each quantum particle.  For two spinless particles, whose joint Hilbert space in the tensor product of two infinite-dimensional position Hilbert spaces, the number of distinct fluids is infinite.  This means that trying to work in this representation will generally be intractable, but it nevertheless provides a clear ontological picture in which all quantum particles can be fully and correctly represented in 3-space.  Importantly, in this ontology, nothing lives in the $3N$-dimensional configuration space, and no physics occurs there.  Instead, the dimension of the joint Hilbert space determines the set of local labels carried by the fluid particle of each individual quantum particle.  The standard quantum formalism is obtained as a delocalizing approximation of this picture, where all of the local states in a particular Lorentz frame are projected onto the time-axis of space-time.

In this new ontological picture, one quantum particle is generally represented by infinitely many classical fluids, and differently-indexed fluid generally only interact during an interaction with another quantum particle, so in the absence of such interaction, each distinct fluid evolves according to the single-particle Schr\"{o}dinger equation, and has its own distinct quantum potential associated with its density.  This follows straightforwardly from the standard $N$-particle Schr\"{o}dinger equation when there is no interaction potential between the particles - even if they are already entangled.  To see this, consider the Schr\"{o}dinger equation written in the following form,
\begin{equation}
    i \hbar \frac{\partial \Psi(\vec{x}_1,\vec{x}_2,\ldots,\vec{x}_N,t)}{\partial t} - \hat{H}\Psi(\vec{x}_1,\vec{x}_2,\ldots,\vec{x}_N,t) = 0
\end{equation}
If there is no interaction potential, then we have $\hat{H} = \hat{H}_1 + \hat{H}_2 + \ldots + \hat{H}_N$, where each $\hat{H}_i$ acts on the spatial degrees of freedom of just one quantum particle.  We can also represent a general entangled wavefunction using the Schmidt decomposition,
\begin{equation}
    \Psi(\vec{x}_1, \vec{x}_2, \ldots,\vec{x}_N,t) =\sum_i \psi_i^1(\vec{x}_1,t)\psi_i^2(\vec{x}_2,t)\ldots\psi_i^N(\vec{x}_N,t).
\end{equation}
Substituting these into the Schr\"{o}dinger equation gives us
\begin{equation}
    i \hbar \sum_i \bigg(  \Big( \frac{\partial \psi_i^1(\vec{x}_1,t)}{\partial t}\psi_i^2(\vec{x}_2,t)\ldots\psi_i^N(\vec{x}_N,t)\Big)   + \Big(\psi_i^1(\vec{x}_1,t)  \frac{\partial \psi_i^2(\vec{x}_2,t)}{\partial t}\ldots\psi_i^N(\vec{x}_N,t)\Big) +\ldots \bigg) \nonumber
\end{equation}
\begin{equation}
    -\Big(\hat{H}_1 +\hat{H}_2 +\ldots\Big) \sum_i \psi_i^1(\vec{x}_1,t)\psi_i^2(\vec{x}_2,t)\ldots\psi_i^N(\vec{x}_N,t) = 0.
\end{equation}
Reorganizing some terms, we have,
\begin{equation}
    0 = \sum_i \bigg(\Big( i\hbar \frac{\partial \psi_i^1(\vec{x}_1,t)}{\partial t} - \hat{H}_1\psi_i^1(\vec{x}_1,t)\Big)\Big(\psi_i^2(\vec{x}_2,t)\ldots\psi_i^N(\vec{x}_N,t)\Big)
\end{equation}
\begin{equation}
    +\psi_i^1(\vec{x}_1,t)\Big( i\hbar \frac{\partial \psi_i^2(\vec{x}_2,t)}{\partial t} - \hat{H}_2\psi_i^2(\vec{x}_2,t)\Big)\Big(\psi_i^3(\vec{x}_3,t)\ldots\psi_i^N(\vec{x}_N,t)\Big) +\ldots\bigg), \nonumber
\end{equation}
from which it clearly follows that
\begin{equation}
    i\hbar \frac{\partial \psi_i^n(\vec{x}_n,t)}{\partial t} - \hat{H}_n\psi_i^n(\vec{x}_n,t) = 0,
\end{equation}
and thus we see that each distinct fluid (index $i$) for each distinct quantum particle (index $n$) evolves according to the single-particle Schr\"{o}dinger equation, as promised - even in standard quantum mechanics.

This motivates the main analysis of this article, where each of these individual fluids has only three spatial dimensions, and so we can let all of them coexist in normal 3-space ($\vec{x}_n \rightarrow \vec{x}$), and each one can then be interpreted using the single-particle Madelung picture.  It is also straightforward to see how this analysis can be extended to include internal degrees of freedom like spin, provided those degrees of freedom are not interacting with the spatial ones, or with each other.

The case of local interaction potentials are represented as local boundary conditions that can mix up the different fluids of each quantum system, change the number of distinct fluids that contain nonzero density and create or destroy entanglement correlations.  This gives rise to a piece-wise wave-field comprising different sets of distinct fluids in different regions, for each quantum particle.  This analysis is too extensive to repeat here, so we refer the reader to \cite{waegell2023local} for more details.

It may be that in practice, local interaction potentials are ubiquitous throughout space, but combining the many single-particle fluid equations with local boundary conditions that connect the sets of fluids on either side of the interaction still ultimately describes all of the dynamics as occurring in 3-space, even if the math is hard to work with.  In the local fluid model, there are only point-particles on world-lines in spacetime, each carrying a local internal information packet with records of past interactions.  There are no nonseparable states describing space-like separated objects in this model, and these are not needed to explain the observable predictions of quantum mechanics.

\end{document}